\documentclass[showpacs,twocolumn,amssymb,prl,aps]{revtex4}

\usepackage{graphicx,epsfig}
\usepackage{dcolumn}
\usepackage{bm}
\begin{document}

\title{Coherent transport on Apollonian networks and continuous-time quantum walks}
\author{Xin-Ping Xu$^{1,3}$}
\author{Wei Li$^{1,2}$}
\author{Feng Liu$^1$}
\affiliation{%
$^1$Institute of Particle Physics, HuaZhong Normal University, Wuhan
430079, China  \\
$^2$Max-Planck-Institute for Mathematics in the Sciences, Inselstr.
22, Leipzig, Germany \\
$^3$Institute of High Energy Physics, Chinese Academy of Science,
Beijing 100049, China
}%

\begin{abstract}
We study the coherent exciton transport on Apollonian networks
generated by simple iterative rules. The coherent exciton dynamics
is modeled by continuous-time quantum walks and we calculate the
transition probabilities between two nodes of the networks. We find
that the transport depends on the initial nodes of the excitation.
For networks less than the second generation the coherent transport
shows perfect revivals when the initial excitation starts at the
central node. For networks of higher generation, the transport only
shows partial revivals. Moreover, we find that the excitation is
most likely to be found at the initial nodes while the coherent
transport to other nodes has a very low probability. In the long
time limit, the transition probabilities show characteristic
patterns with identical values of limiting probabilities. Finally,
the dynamics of quantum transport are compared with the classical
transport modeled by continuous-time random walks.
\end{abstract}
 \maketitle
The problem of coherent exciton transport modeled by quantum walks
is widely studied and relevant to many distinct fields, such as
polymer physics, solid state physics, biological physics and quantum
computation~\cite{rn1,rn2,rn3}. Such studies have been done in the
framework of continuous-time quantum walks (CTQWs) and on various
discrete systems~\cite{rn4,rn5}. It has been shown that the dynamics
of coherent transport are strongly influenced by the structure of
the underlying discrete systems~\cite{rn6,rn7,rn8}. Most of previous
studies focus CTQWs on graphs with simple
structures~\cite{rn9,rn10,rn11}, coherent dynamics on general graphs
have not received much attention in the scientific community. To
this end, it is natural to consider quantum transport on graphs with
general structure embedded in nature.

An important and universal feature of networked systems (or graphs)
in nature is that they have the small-world and scale-free
property~\cite{rn12,rn13}. The Apollonian networks
(ANs)~\cite{rn14,rn15} are a very useful toy model that captures all
these features simultaneously, thus provide a good facility to study
the dynamical processes taking place on networked systems, including
percolation, electrical conduction, etc~\cite{rn14,rn16,rn17}.

\begin{figure}
\scalebox{0.8}[0.8]{\includegraphics{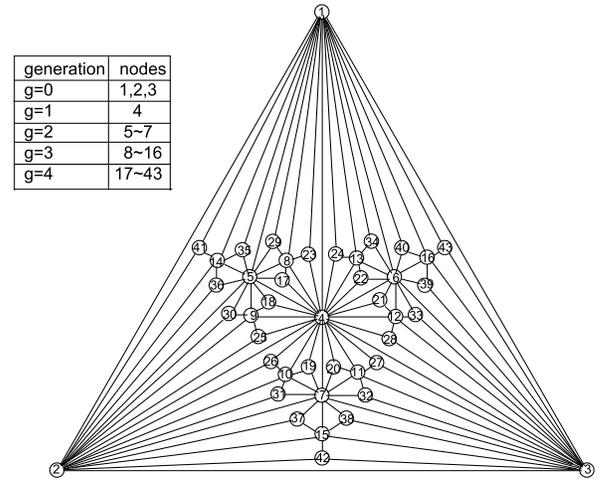}} \caption{
Apollonian network generated by simple iterative rules in four
generations ($G=4$). The nodes at each generation $g$ are marked as
consecutive numbers.
 \label{fg1}}
\end{figure}

In this paper, we consider coherent exciton transport on 2D
Apollonian networks (ANs). The network can be generated as
follows~\cite{rn14}: At the initial generation $g=0$, the network is
composed of three fully connected nodes marked as $1$, $2$, and $3$.
At the subsequent generation, a new node is added inside each (newly
established) triangle and linked to the three vertices of the
triangle. Using this simple rule, we can obtain a deterministic 2D
ANs of size $N=3+(3^G-1)/2$ ($G$ is the number of
generation)~\cite{rn14}. Many topological properties of this network
model have been well-studied in the literature~\cite{rn15,rn18}.
Fig.~\ref{fg1} shows the structure of an AN in four generations
($G=4$).
\begin{figure}
\scalebox{0.7}[0.56]{\includegraphics{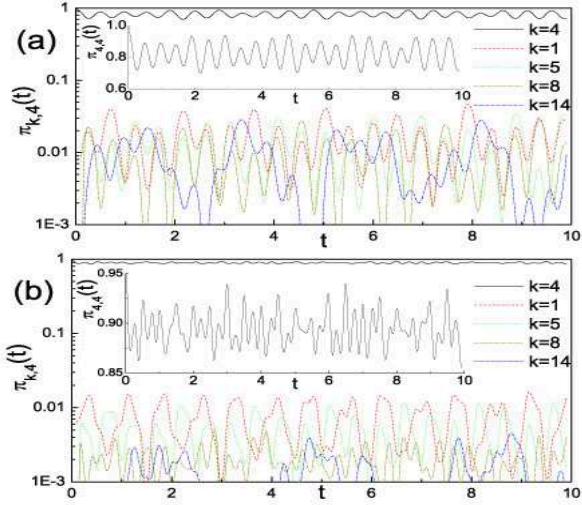}} \caption{(Color
online) Time evolution of transition probabilities $\pi_{k,4}(t)$
for different values of $k$ (marked as different types of curves) on
ANs of $G=3$ (a) and $G=4$ (b). The excitation starts at central
node $4$. The insets are enlarged linear-scale plots of return
probability $\pi_{4,4}(t)$.
 \label{fg2}}
\end{figure}

The coherent exciton transport on a connected network is modeled by
the continuous-time quantum walks (CTQWs), which is obtained by
replacing the Hamiltonian of the system by the classical transfer
matrix, i.e., $H=-T$~\cite{rn19,rn20}. The transfer matrix $T$
relates to the Laplace matrix by $T=-\gamma A$, where for simplicity
we assume the transmission rates $\gamma$ of all bonds to be equal
and set $\gamma \equiv 1$ in the following~\cite{rn19,rn20}. The
Laplace matrix $A$ has nondiagonal elements $A_{ij}$ equal to $-1$
if nodes $i$ and $j$ are connected and $0$ otherwise. The diagonal
elements $A_{ii}$ equal to degree of node $i$, i.e., $A_{ii}=k_i$.
The states $|j\rangle$ endowed with the node $j$ of the network form
a complete, ortho-normalised basis set, which span the whole
accessible Hilbert space. The time evolution of a state $|j\rangle$
starting at time $t_0$ is given by $|j,t\rangle =
U(t,t_0)|j\rangle$, where $U(t,t_0)=exp[-iH(t-t_0)]$ is the quantum
mechanical time evolution operator. The transition amplitude
$\alpha_{k,j}(t)$ from state $|j\rangle$ at time $0$ to state
$|k\rangle$ at time $t$ reads $\alpha_{k,j}(t)=\langle
k|U(t,0)|j\rangle$ and obeys Schr\"{o}dinger¡¯s
equation~\cite{rn21}. Then the classical and quantum transition
probabilities to go from the state $|j\rangle$ at time $0$ to the
state $|k\rangle$ at time $t$ are given by $p_{k,j}(t)=\langle
k|e^{-tA}|j\rangle$ and $\pi_{k,j}(t)=|\alpha_{k,j}(t)|^2= |\langle
k|e^{-itH}|j\rangle|^2$~\cite{rn19}, respectively. Using $E_n$ and
$|q_n\rangle$ to represent the $n$th eigenvalue and eigenvector of
$H$, the classical and quantum transition probabilities between two
nodes can be written as~\cite{rn19,rn20,rn21}
\begin{equation}\label{eq1}
p_{k,j}(t)=\sum_n e^{-tE_n}\langle k|q_n\rangle \langle
q_n|j\rangle,
\end{equation}
\begin{equation}\label{eq2}
\pi_{k,j}(t)=|\alpha_{k,j}(t)|^2=|\sum_n e^{-itE_n}\langle
k|q_n\rangle \langle q_n|j\rangle|^2.
\end{equation} Generally, to
get $p_{k,j}(t)$ and $\pi_{k,j}(t)$, all the eigenvalues $E_n$ and
eigenvectors $|q_n\rangle$ are required. In the following we will
consider $p_{k,j}(t)$ and $\pi_{k,j}(t)$ obtained from diagonalizing
the Hamiltonian $H$ by using the standard software package
Mathematica 5.0.
\begin{figure}
\scalebox{0.7}[0.6]{\includegraphics{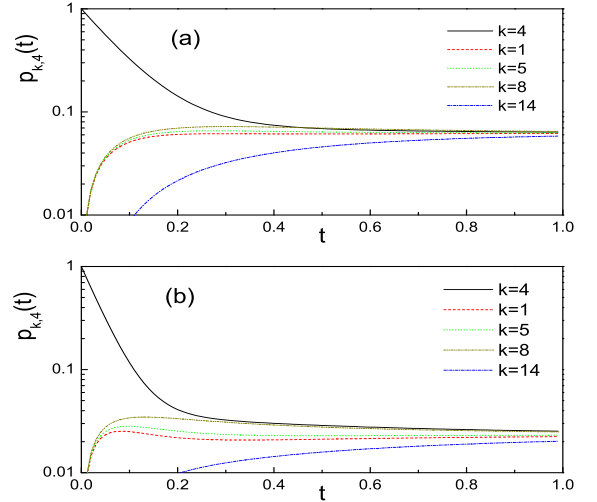}}
 \caption{(Color online)
Time evolution of the classical probabilities $p_{k,4}(t)$ for
different values of $k$ (marked as different types of curves) on ANs
of $G=3$ (a) and $G=4$ (b). The excitation starts at central node
$4$. The classical $p_{k,4}(t)$ approach the equip-partitioned
probability $1/N$ at long time scale.
 \label{fg3}}
\end{figure}

We start our analysis by considering transport dynamics on ANs of
$G=3$ ($N=16$) and $G=4$ ($N=43$) when the excitation starts at the
central node $4$. The nodes are numbered according to Fig.~\ref{fg1}
and network of $G=3$ (nodes labeled as $1\sim16$) is a subgraph of
$G=4$ (nodes labeled as $1\sim43$). Due to rotational symmetry, the
transition probabilities from node $4$ to certain groups of nodes
are equal. Thus, we choose several different transition
probabilities, namely, $\pi_{4,4}(t)$, $\pi_{1,4}(t)$,
$\pi_{5,4}(t)$, $\pi_{8,4}(t)$ and $\pi_{14,4}(t)$ for further
study.

Fig.~\ref{fg2} shows these quantum transition probabilities for ANs
of $G=3$ and $G=4$. We find that there is a high probability to find
the excitation at the initial node ($\pi_{4,4}(t)$ marked as solid
curves in the Fig.~\ref{fg2}). To see the behavior of $\pi_{4,4}(t)$
clearly, we display $\pi_{4,4}(t)$ in an enlarged linear scale (See
inserted plots in Fig.~\ref{fg2}). For AN of $G=3$, $\pi_{4,4}(t)$
shows regular oscillations, as generation increases, $\pi_{4,4}(t)$
becomes irregular and its average value increases (Compare the
inserted plots in Fig.~\ref{fg2} (a) and (b)). Transition
probabilities between the initial node and other nodes are
considerably low compared to the return probability $\pi_{4,4}(t)$
(See the dashed curves in (a) and (b)). The corresponding transition
probabilities $\pi_{k,4}(t)$ ($k\neq 4$) of a $G=3$ AN is higher
than those of a $G=4$ AN. This may be attributed to the fact that
the return probability $\pi_{4,4}(t)$ on a $G=4$ AN is larger than
that on a $G=3$ AN.
\begin{figure}
\scalebox{0.7}[0.5]{\includegraphics{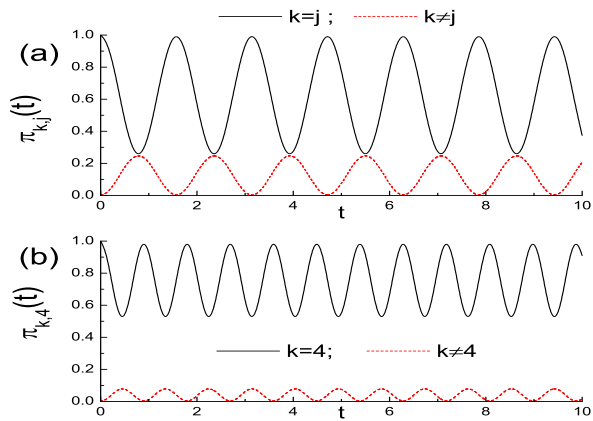}} \caption{(Color
online) (a) Transition probabilities $\pi_{k,j}(t)$ for AN of $G=1$.
(b) Transition probabilities $\pi_{k,4}(t)$ for AN of $G=2$. Both
results are numerically obtained by diagonalizing the Hamiltonian
$H$ and consistent with the analytical results in Eqs.~(\ref{eq3})
and (\ref{eq4}).
 \label{fg4}}
\end{figure}

Fig.~\ref{fg3} shows the classical transition probabilities
$p_{k,4}(t)$ for different values of $k$. It is found that the
classical transition probabilities approach the equipartition $1/N$
very quickly and $p_{14,4}(t)$ reaches $1/N$ much slower than other
transition probabilities. This can be explained by the shortest path
length from the initial excitation node $4$. The number of bonds
between node $4$ and $14$ is larger than the distance between other
pairs of nodes. In addition, because the shortest path lengths
between $4$ and $1$, $4$ and $5$, $4$ and $8$ are equal, the
classical $p_{1,4}(t)$, $p_{5,4}(t)$ and $p_{8,4}(t)$ are comparable
(Compare the curves in Fig.~\ref{fg3}). Noting that the long time
averaged $\pi_{4,4}(t)$ is much higher than equip-partitioned
probability $1/N$ and other (long time averaged) quantum transition
probabilities is less than $1/N$, we conclude that the classical
transport to other nodes (non-initial node) of the network is more
efficient than quantum transport.

Interestingly, for $G=1$ ($N=4$) and $G=2$ ($N=7$) ANs, the quantum
transition probabilities are fully periodic when the coherent
excitation starts from the central node $4$. In this case we obtain,
based on the analytically determined eigenvalues and
eigenvectors~\cite{rn22}, that for $G=1$, $\pi_{k,j}(t)$ have the
following, periodic form:
\begin{equation}\label{eq3}
 \pi_{k,j}(t)=\left\{
\begin{array}{ll}
(5+3\cos4t)/8,   &  k=j, \\
(1-\cos4t)/8,  & k \neq j.
\end{array}
\right.
\end{equation}
And for $G=2$ we have,
\begin{equation}\label{eq4}
 \pi_{k,4}(t)=\left\{
\begin{array}{ll}
(37+12\cos7t)/49,   &  k=4, \\
(2-2\cos7t)/49,  & k \neq 4.
\end{array}
\right.
\end{equation}
Fig.~\ref{fg4} shows the behavior of $\pi_{k,j}(t)$ obtained by
numerically diagonalizing the Hamiltonian $H$ for $G=1$ and $G=2$
ANs. This agrees the analytical results in Eqs.~(\ref{eq3}) and
(\ref{eq4}). We find that there is a perfect revival of the initial
state for each $t=2n\pi /N$ ($n\in$ Integers), where $N$ is the
number of nodes of the considered network. This revival of the
initial probability distribution resembles the results obtained for
continuous and discrete quantum carpets~\cite{rn23,rn7}, in which
the revival is only perfect for small size of
cycles~\cite{rn24,rn7}. The case for ANs is analogous: The revivals
are perfect for small ANs of $G\leqslant 2$, when the network size
becomes larger ($G\geqslant 3$), there are only partial revivals of
the initial state (Compare Fig.~\ref{fg2} and Fig.~\ref{fg4}).
\begin{figure}
\scalebox{0.7}[0.5]{\includegraphics{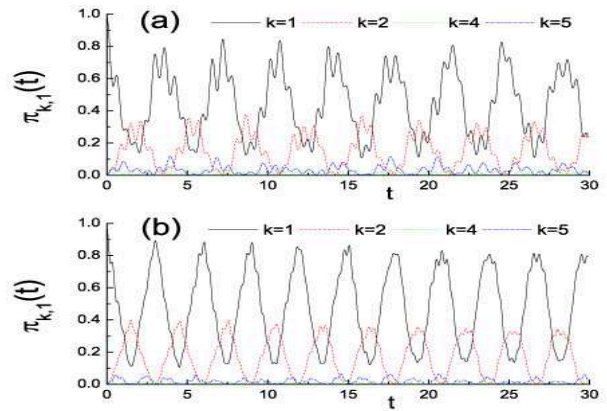}} \caption{ (Color
online) Time evolution of transition probabilities $\pi_{k,1}(t)$
for different values of $k$ (marked as different types of curves) on
ANs of $G=3$ (a) and $G=4$ (b). The excitation starts at noncentral
node $1$. The return probability $\pi_{1,1}(t)$ is nearly periodic
for both the networks. \label{fg5} }
\end{figure}

Now we turn to the case when the initial excitation starts at other
positions of the network. Fig.~\ref{fg5} shows the transition
probabilities when the initial excitation is placed at node $1$. For
both the $G=3$ and $G=4$ ANs, the return probabilities
$\pi_{1,1}(t)$ display regular oscillations. The return probability
$\pi_{1,1}(t)$ is much larger than other transition probabilities
$\pi_{k,1}(t)$ ($k\neq 1$) at most time intervals. It is interesting
to note that except for the high return probability $\pi_{1,1}(t)$,
there is also considerable transport to nodes $2$ and $3$ (Note
$\pi_{2,1}(t)=\pi_{3,1}(t)$ because of axis-symmetry). Nevertheless,
transport to other nodes (such as $4$, $5$ etc) is particularly low.
This suggests that the excitation is preferably located on the nodes
of the same generation of the initial node.

If the initial excitation starts from other noncentral nodes, the
results are similar but some details change. The oscillation
amplitude and period are different and there is also a relative high
probability to find the excitation at the initial node.
\begin{figure}
\scalebox{0.7}[0.8]{\includegraphics{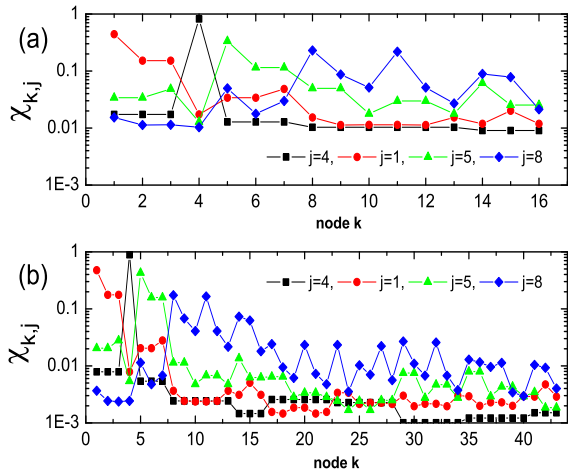}} \caption{ (Color
online) Long time limiting probabilities $\chi_{k,j}$ for different
node $j$ of initial excitation on the $G=3$ (a) and $G=4$ (b) ANs.
The squares, dots, triangles and rhombus denote initial excitation
at node $4$, $1$, $5$ and $8$ respectively.
 \label{fg6} }
\end{figure}

In order to discuss what happens at long times, we consider the long
time averages of the transition probabilities $p_{j,k}(t)$ and
$\pi_{j,k}(t)$. On finite ANs, the transition probability converges
to a certain value, this value is determined by the long time
average. Classically, the long time averaged transition
probabilities equal to the equal-partitioned probability $1/N$.
However, the quantum transport does not lead to equipartition. The
long time average of $\pi_{j,k}(t)$ is defined as
\begin{equation}\label{eq5}
\begin{array}{ll}
\chi_{k,j}&=\lim_{T\rightarrow \infty}\frac{1}{T}\int_0^T
\pi_{k,j}(t)dt \\
&=\sum_{n,l}\delta (E_n-E_l)\langle k|q_n\rangle \langle
q_n|j\rangle \langle j|q_l\rangle \langle q_l|k\rangle,
\end{array}
\end{equation}
where $\delta (E_n-E_l)=1$ for $E_n=E_l$ and $\delta (E_n-E_l)=0$
else. Some eigenvalues of $H$ may be degenerate, so the sum in the
equation contain terms belonging to different eigenstates. Here, we
consider the limiting transition probabilities $\chi_{k,j}$
according to this equation.

Fig.~\ref{fg6} (a) and (b) show the limiting probability
distributions for $G=3$ and $G=4$ ANs. In the figure, we find that
$\chi_{j,j}$ is larger than other transition probabilities
$\chi_{k,j}$ ($k\neq j$). This indicates the excitation is most
likely to be found at the initial node, which is accord with the
observation in Figs.~\ref{fg2} and \ref{fg5}.

An interesting feature related to the limiting probabilities is that
different nodes, $k$ and $l$, may have the same transition
probabilities, i.e., $\chi_{k,j}=\chi_{l,j}$. Concretely, for an
excitation starting from the central node $j=4$, transport to nodes
of certain cluster has identical limiting transition probabilities
(See the black squares in Fig.~\ref{fg6}). For instance,
$\chi_{1,4}$, $\chi_{2,4}$ and $\chi_{3,4}$ are equal to each other;
$\chi_{5,4}=\chi_{6,4}=\chi_{7,4}$; $\chi_{k,4}$ are equal for
$8\leqslant k\leqslant 13$; ... . The nodes of clusters having the
same transition probabilities in such case are easy to be identified
due to rotation-symmetry of the central node. Furthermore, nodes of
the same generation or having the same connectivity may have
different limiting probabilities (compare the values of $\chi_{k,4}$
for cluster $8\leqslant k\leqslant 13$ and cluster $14\leqslant
k\leqslant 16$ in the Fig.~\ref{fg6}).

For an excitation starting from noncentral node, the situation is
quite different. When the excitation starts at node $1$ (See the
dots in Fig~\ref{fg6}), $\chi_{2,1}$ equals to $\chi_{3,1}$. Such
identical values of transition probabilities are also easy to be
distinguished and can be understood as a result of the
axis-symmetry. The case for excitation starting at node $5$ is
analogous (See the triangles in the plots). Particularly, if the
excitation starts at node $8$, nodes $10$ and $12$ have the same
limiting probability, i.e., $\chi_{10,8}=\chi_{12,8}$ (rhombus
indicated in Fig.~\ref{fg6}). Such kind of identical probability is
not straightforward to be realized but also can be ascribed to the
rotation symmetry of the structure of ANs. If triangle $\Delta 145$
is rotated $\pi/3$ and $2\pi/3$, the initial node $8$ changes to the
positions $12$ and $10$ respectively. Except for the equal value of
$\chi_{10,8}$ and $\chi_{12,8}$ on both the $G=3$ and $G=4$ ANs, we
find that there are more identical probabilities on the $G=4$ AN.
For instance, we find the following equal transition probabilities:
$\chi_{31,8}=\chi_{33,8}$, $\chi_{19,17}=\chi_{21,17}$,
$\chi_{26,23}=\chi_{28,23}$, $\chi_{10,29}=\chi_{12,29}$,
$\chi_{31,29}=\chi_{33,29}$, $\chi_{37,35}=\chi_{39,35}$, etc. For
ANs of higher generation, the previous identical limiting
probabilities are preserved and additional identical values of
transition probabilities are observed due to structural symmetry of
the network.

In summary, we have studied coherent exciton transport modeled by
continuous-time quantum walks on ANs. The quantum transport exhibits
a very distinct behavior compared to the classical random walks. For
networks up to the second generation the coherent transport shows
perfect recurrences when the initial excitation starts at the
central node~\cite{rn7}. For networks of higher generation, the
recurrence ceases to be perfect, which resembles results for
discrete quantum carpets~\cite{rn7}. The excitation depends on the
initial nodes and is most likely to be found at the original nodes
while the coherent transport to other nodes is particularly low. In
the long time limit, the transition probabilities show identical
values between different nodes, which reflects the symmetry of the
network structure.

We would like to point out that although CTQWs on ANs show
oscillation and revivals like the results of the 1D case, there are
some difference in the quantum dynamics between the two structures.
For ANs, we find that the return probabilities at the central nodes
are nearly periodic, in contrast to the 1D case where the (maximums
of) return probability shows a power law decay as $\pi(t) \sim
t^{-1}$~\cite{rn21,rn25}. In Ref.~\cite{rn25}, the authors find that
for a 1D chain, quantum revivals do not repeat indefinitely but
become less and less accurate as time progresses~\cite{rn25}. They
also find that the quantum walks displays Anderson localizations or
decoherence in the presence static or dynamic disorder~\cite{rn25}.
For ANs, there are also considerable localizations on the initial
nodes (See Fig.~\ref{fg6}). Such localizations may relate to the
network structures and requires a further study.

This work is supported by National Natural Science Foundation of
China under Project Nos 10575042, 10775058 and MOE of China under
contract number IRT0624 (CCNU).

\end{document}